\date{}
\documentstyle[aps,epsf]{revtex}
\tightenlines
\begin{document}
\title
{Persistent currents  in mesoscopic cavities with surface roughness}
\author{ V. M. Apel, M. J. S\'anchez and G. Chiappe}
\maketitle
\noindent
\begin{center}
\center{\it Departamento de F\'{\i}sica  J. J. Giambiagi,\\
 Facultad de Ciencias Exactas  y Naturales, \\
Universidad de Buenos Aires.
Ciudad Universitaria, 1428 Buenos Aires, Argentina.}\\ 
\end{center}
\begin{abstract}
We consider  a bidimensional discrete annular cavity with surface roughness (SR) threaded by a magnetic flux. In the ballistic regime and at half filling, localized border-states show up.
These border states contribute coherentely to the persistent current and  the magnitude of the typical current $I_{typ}$ is enhaced with respect to their value in the absence of confinement.
This result is robust with respect to changes in the relevant lengths of the sample.
\end{abstract}
\vspace{2cm}

In recent years, advances in  nanotechnology  made possible  to design mesoscopic samples  in which the carriers are mainly scattered by the boundaries of the system \cite{kouwen}. 
In this situation, where the elastic mean free path is much larger than the system size, the sample can be considered in the ballistic regime in order to compute the relevant transport quantities.
In addition, the potential used to confine the carriers into the mesoscopic device has a symmetry  quite different from that of the lattice of the sample.
Therefore, even when the bulk disorder is absent, surface roughness (SR) is present with the consequence of the irregular scattering of the carriers
at the borders of the sample.

In this article we present a model of SR which emerges naturally
when the topology of the discrete lattice and the confining potential are different. 
The relevant feature of our model is that the effect  of the SR on the eigenstates depends on the  energy of these states. As the energy spectrum is bounded,  the lattice parameter ${a}$
is a  cut off for the allowed wavelengths and the  wavelenghts close to  ${a}$ will be the most sensitives to the SR. 
Therefore, the SR that we model will be relevant near half filling, {\it i.e} in the {\it metallic regime}.

When the sample is threaded by a magnetic flux persistent currents are originated \cite{but,cheung1}. Although much work has been done concerning the influence of bulk disorder on persistent currents \cite{guhr},
less is known about the ballistic regime. Our goal is to characterize the effect of the confinement and the SR on the magnitude of the persistent current in that regime.

The total current   can be calculated at zero temperature as 
$I \sim {\partial E /\partial \phi}$, with $E$ the total energy of the system and $\phi$ the magnetic flux.  
The value of $I_{typ} \equiv \sqrt{\int I^2 \; d\phi}$  depends on the number of open channels present in the system $M$. For discrete sytems with cylindrical geometry  and away from half filling, $I_{typ} \sim \sqrt{M}$ \cite{cheung2,bouch}.
This behavior can be understood taking into account the properties of the crossings (or quasicrossings) that appear  in the spectrum when the flux is varied \cite{nakamura}. They  occur mainly between levels belonging to different channels and frustrate the coherent increased of the current \cite{fs}.
Although very particular geometric configurations can give rise to coherent contributions of all the channels \cite{cheung2,louis}, those sharp configurations are very difficult to obtain experimentally.  

We consider a $N \times N$ cluster on a square lattice. In the following all the lengths will be given in units of ${a}$.
Taking a single  atomic level per  lattice site the Hamiltonian is,
\begin{equation}
H= \sum_{m,n} \epsilon_{m,n}|m,n><m,n|+
\sum_{m,n;m',n'} t_{m,n;m',n'}|m,n><m',n'| \; , 
\end{equation}
where $(m,n) \equiv {\bf m}$ labels the coordinates of the sites in the lattice.
The hopping integrals
$t_{\bf{m};\bf{m'}}$ are restricted to nearest neighbours. Assuming that
the vector potential ${\bf A}$ has only azimutal component we take
$t_{\bf{m};\bf{m+l}} = t \; \exp{( i \int_{\bf{m}}^{\bf{m+l}} \bf{A} {\bf{.}} d \bf{l} )} \;$,
in units of the quantum flux $\phi_{o}$, where  $\bf{l}$ is a vector that points from the site $\bf m$ to any of its four nearest neighbours.\\
With the on-site energies, $\epsilon_{\bf m}$, 
we simulate the confining potential in order to obtain the required profile of the sample.  In particular for an annular cavity with internal radius $r$ and external radius $R$, we define
\begin{equation}
\epsilon_{\bf m} = \left\{ \begin{array}{lll}
                   0   & \; \; \; \; $if$ \; r \le  |m | \le R \;\\
		   V   & \; \; \; \; \mbox{otherwise} \; ,	
				\end{array}
                 \right.  \nonumber 
\end{equation}
where $V$ is large enough  to confine the carriers within the annulus.
For the present  geometry the two  relevant  lenght scales can be chosen as the internal  radius  $r$ and the mean  width of the sample $W$. While the second determines $M$ for a given value of the Fermi energy \cite{canal}, the ratio 
$s \equiv W / r$ controls the strength of SR. For  $s >> 1$ and for carriers with a wavelength of the order of $a$, two succesive scattering events by  the inner and outer boundaries of the sample are uncorrelated. 
On the other hand, for $s <<1$ and the same wavelengths those  events are strongly correlated. Therefore the eigenstates corresponding to these particular wavelenghts will evolve from generic extended ones to  strongly localized as the value of $s$ is reduced. 

In order to describe our results we will focus in two representative samples:
S1, with $W= 40$  and $r= 20$ and 
S2, with $W= 16$ and  $r= 90$. The values are given in units of $a$. In both samples the averaged area of the sample  is $A$ and equal to the total number of sites. 

\begin{figure}
\begin{center}
\leavevmode
\epsfysize=10cm
\epsfbox[12 12 559 756]{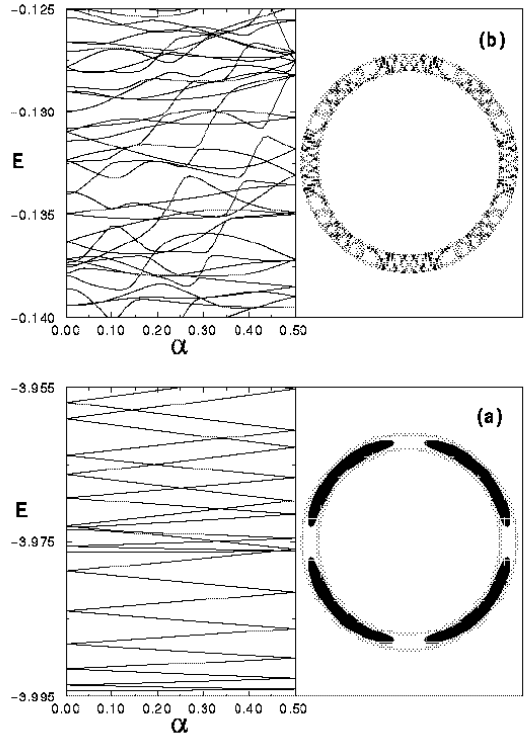}
\end{center}
\caption{(a) Energy levels at the bottom of the spectrum  as a function of the  rescaled magnetic flux $\alpha$ for sample S1 together with a typical charge distribution. (b) Idem as (a) but for the quarter filling region of the spectrum.} 
\label{1}
\end{figure}

In Fig.~\ref{1} we show for sample $S1$ the bottom (Fig.~\ref{1}(a)) and the quarter filling (Fig.~\ref{1}(b))  regions of the energy spectrum  as a function of the normalized flux $\alpha = \phi / \phi_{0}$ together with a
generic  charge distribution for each region.
Fig.~\ref{1}(a) corresponds to  the large wavelength numbers, that are the less sensitive ones to the effect of the SR. This region of the spectrum looks qualitative similar to the spectrum of an integrable Aharonov-Bohm annular billiard  \cite{fs}. 
That is,  crossings  between levels belonging to the same channel at  $\alpha = 0$ and $\alpha = 1 /2$ and crossings   between levels belonging to different channels for $0< \alpha < 1/2 $. Moreover, a generic  charge distribution corresponds to eigenstates with well defined value of the angular quantum number.

\begin{figure}
\begin{center}
\leavevmode
\epsfysize=12cm
\epsfbox[12 12 559 756]{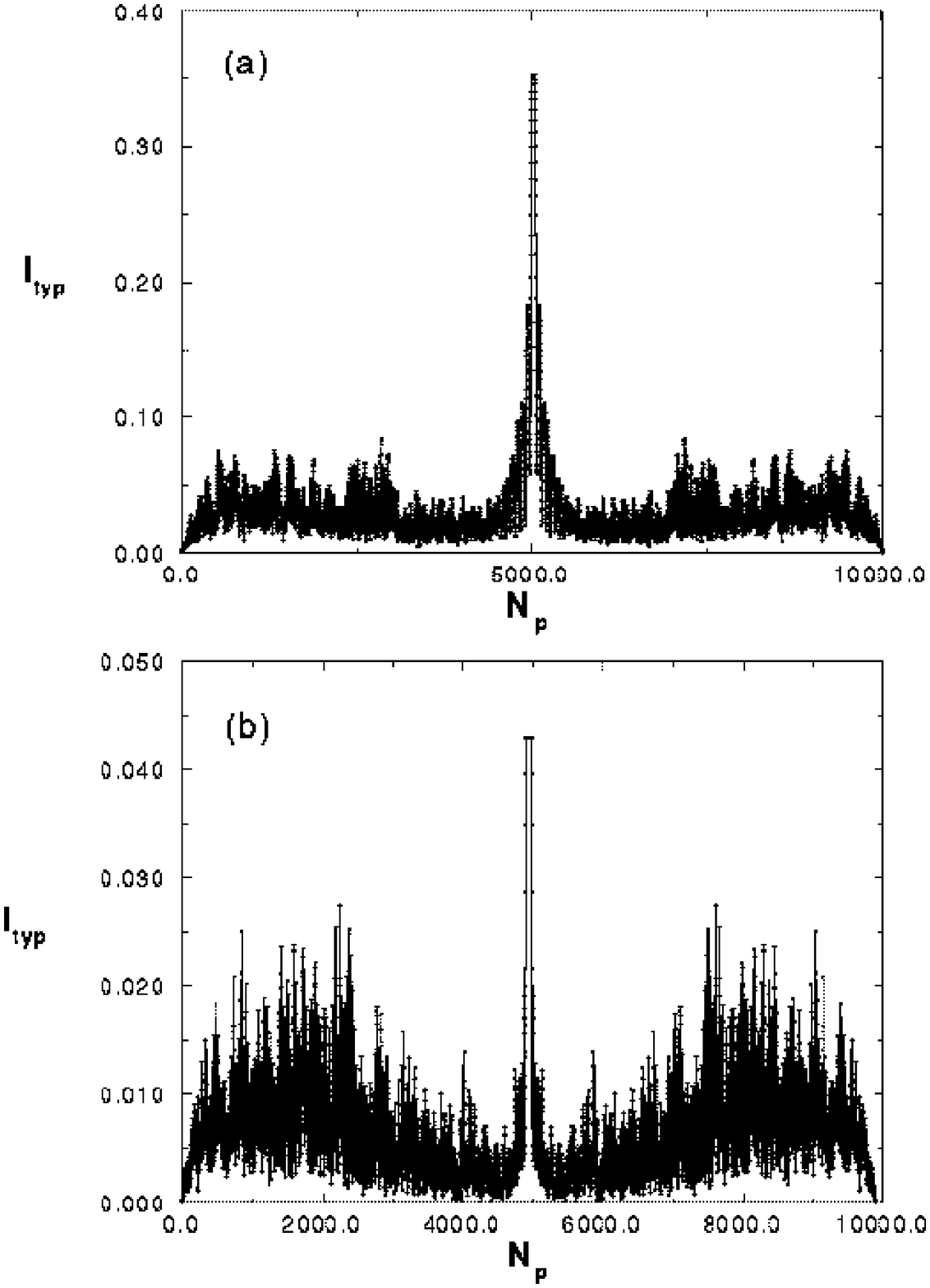}
\end{center}
\caption{(a) Typical current as a function of $N_{p}$ for sample $S1$.(b)
Idem as (a) for sample $S2$.} 
\label{3}
\end{figure}

In the quarter filling region of the spectrum   the typical wavelengths are greater than ${a}$ but small  enough to make  the states sensitive to the global  polygonal shape of the sample. Therefore, as 
the energy increases,  avoided crossings are showed up as a function of $\alpha$.
The corresponding charge distributions are quite similar to those  of  a continuous polygonal billiard and are extended states in general (see  Fig.~{1}(b)). 
In both  energy regions, when an occuppied level crosses (or quasicrosses)  an empty one, the total current $I$ exhibits a discontinuity (or abrupt oscillation). The magnitude of $I_{typ}$ will be reduced 
with respect to the situation with no crossings  in the interval $(0, \alpha /2)$  \cite{cheung2,louis}.
As it is shown in Fig.~\ref{3}, below half filling, $I_{typ}$ is a highly fluctuating function of the number of particles  $N_{p}$. This is a consequence of the randomness in the  distribution of crossings (or quasicrossings) present in these regions of the spectrum \cite{fs}.
For sample $S2$ a qualitative similar behavior is found.

\begin{figure}
\begin{center}
\leavevmode
\epsfysize=12cm
\epsfbox[12 12 559 756]{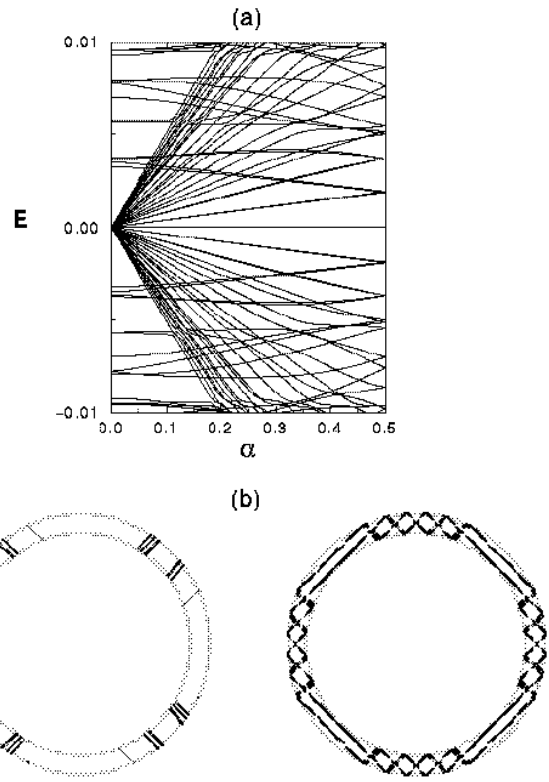}
\end{center}
\caption{(a) Energy levels   close to half filling as a function of the rescaled magnetic flux $\alpha$ for $S1$.(b) Typical charge distributions for angular localized states (left side) and radially localized states (right side). } 
\label{2}
\end{figure}

The upper region of the spectrum, close to half filling, is shown in Fig.~\ref{2}(a) for $S1$.
As we have already mentioned, in this region  the states are sensitive to the fine structure of the boundaries of the sample. A bunching  of quasi-degenerated eigenenergies appears at zero flux corresponding to two different types of  localized states.
One type are the  angular localized  states whose characteristic charge distribution is depicted in the left side  of Fig.~\ref{2}(b).
These states are trapped  in regions where the sample has more steps on the boundaries. 
They are unsensitive to the magnetic flux and are the origin of the
flat lines in the spectrum.
The other type of states are mostly radially localized and look like whispering gallery modes. These states are sensitive to the magnetic flux and therefore will carry a finite current. The right side of Fig.~\ref{2}(b) displays the characteristic charge distributions  for these states.
We  remark that qualitatively the same description holds for sample $S2$. Nevertheless, as the fraction of angular localized states decreases with the width  $W$ of the sample, the number of  quasidegenerated flat levels in the spectrum of $S2$ at half filling will be greater than in $S1$.

Opposite to what happens for models with diagonal disorder, for the present system the  localization is obtained without introducing an additional scale of energy. Only remains the scale  defined by the kinetic energy. Therefore, at zero flux, the energy of the localized states  will be very close to  zero.
This is a consequence of the finite character of the spectrum and its symmetry. 

In order to quantify the described effect we have to consider two energy scales. One is the mean level spacing $\Delta$ near half filling between levels of a given channel and in the absence of confinement, $ \Delta = \frac{ M}{  \rho}  $,  where $\rho$ is the mean density of states near half filling \cite{cheung2}.
The other relevant energy scale, $ \Delta_{g}$, is a measure of the effect of the confinement on the eigenstates and eigenvalues of the system without borders.
To obtain a generic quantitative relation between  $\Delta_{g}$ and $\Delta$ is not a simple task. However we present the following argument: for 
$W \rightarrow \infty$ the effect of the confinement tends to zero, and then $\Delta_{g} < \Delta / M$. On the other hand, for $W \rightarrow 0$ the effect of the SR is maximized and  $\Delta_{g} > \Delta$. In this regime the eigenstates become angular localized, $\it i.e$ are not current carrying.
By increasing $W$ we can go from one limit to the other one.
Therefore in a wide intermediate regime it should be
satisfied that $ \frac{\Delta}{M} < \Delta_{g} < \Delta $.
When this condition is acomplished the interaction with the borders mixes 
states belonging almost all to different channels ($\it i.e$ mainly with
similar angular momentum), making radially localized eigenfunctions, which corresponds {\it to the  quasidegenerated levels bunching at zero energy that are current carrying}. Therefore even in the presence of SR, typical states like the whispering gallery modes survive in the discrete system close to half filling.
This have an important consequence on the values of $I_{typ}$  close to half filling, as it is shown in Fig.~\ref{3} for both samples $S1$ and $S2$. 
As there are not crossings between occupied and empty levels in all the region of variation of the flux, $I_{typ}$  increases monotonically with $N_{p}$. A plateau occurs at the maximum value of $I_{typ}$, 
and appears when the flat levels (angular localized states) begin to be filled. 
As we have already mentioned the width of the plateau for $S1$ (Fig.~\ref{3}(a))
is smaller than for $S2$ (Fig.~\ref{3}(b)).

As far as we know the enhacement of the persistent currents due to confinement effects in discrete systems has not been adressed in previous theoretical works. We believe that the present work  could  help to 
understand early  experimental resuls in $Au$ rings \cite{chandra}
which have shown  big discrepancies with the existing theories. 
   
\section*{Acknowledgments}
This work was partially supported by UBACYT (TW35 - TW61), PICT97 03-00050-01015 and CONICET. We acknowledge useful discussions with A. J. Fendrik.
\nopagebreak

\newpage

\begin{thebibliography}{99}
\bibitem{canal} As usual the number of channels $M$ is defined as the ratio between the mean width of the sample $W$ and the Fermi wavelength.
\end{thebibliography}
\end{document}